\documentclass[twocolumn]{aastex631}
\usepackage[version=3]{mhchem}
\usepackage{threeparttable, tablefootnote}
\begin{document}

\title{Measuring the \ce{^{34}S} and \ce{^{33}S} isotopic ratios of volatile sulfur during planet formation}
\correspondingauthor{Alice S. Booth} 
\email{alice.booth@cfa.harvard.edu}
\author[0000-0003-2014-2121]{Alice S. Booth} 
\altaffiliation{Clay Postdoctoral Fellow}
\affiliation{Center for Astrophysics \textbar\, Harvard \& Smithsonian, 60 Garden St., Cambridge, MA 02138, USA}

\author[0000-0001-7479-4948]{Maria N. Drozdovskaya}
\affiliation{Physikalisch-Meteorologisches Observatorium Davos und Weltstrahlungszentrum (PMOD/WRC), Dorfstrasse 33, CH-7260, Davos Dorf, Switzerland}

\author[0000-0002-7935-7445]{Milou Temmink}
\affiliation{Leiden Observatory, Leiden University, 2300 RA Leiden, the Netherlands}

\author[0000-0002-7058-7682]{Hideko Nomura}
\affiliation{Division of Science, National Astronomical Observatory of Japan, 2-21-1 Osawa, Mitaka, Tokyo 181-8588, Japan}

% alphabetical order 
\author[0000-0001-7591-1907]{Ewine F. van Dishoeck}
\affiliation{Leiden Observatory, Leiden University, 2300 RA Leiden, the Netherlands}
\affiliation{Max-Planck-Institut für Extraterrestrische Physik, Giessenbachstrasse 1, 85748 Garching, Germany}

\author[0000-0001-5849-577X]{Luke Keyte}
\affiliation{Department of Physics and Astronomy, University College London, Gower Street, London, WC1E 6BT, UK}

\author[0000-0003-1413-1776]{Charles J.\ Law}
\altaffiliation{NASA Hubble Fellowship Program Sagan Fellow}
\affiliation{Department of Astronomy, University of Virginia, Charlottesville, VA 22904, USA}

\author[0000-0003-3674-7512]{Margot Leemker}
\affiliation{Leiden Observatory, Leiden University, 2300 RA Leiden, the Netherlands}

\author[0000-0003-2458-9756]{Nienke van der Marel}
\affiliation{Leiden Observatory, Leiden University, 2300 RA Leiden, the Netherlands}

\author[0000-0003-2493-912X]{Shota Notsu}
\affiliation{Department of Earth and Planetary Science, Graduate School of Science, The University of Tokyo, 7-3-1 Hongo, Bunkyo-ku, Tokyo 113-0033, Japan}
\affiliation{Department of Astronomy, Graduate School of Science, The University of Tokyo, 7-3-1 Hongo, Bunkyo-ku, Tokyo 113-0033, Japan}
\affiliation{Star and Planet Formation Laboratory, RIKEN Cluster for Pioneering Research, 2-1 Hirosawa, Wako, Saitama 351-0198, Japan}

\author[0000-0001-8798-1347]{Karin Öberg}
\affiliation{Center for Astrophysics \textbar\, Harvard \& Smithsonian, 60 Garden St., Cambridge, MA 02138, USA}

\author[0000-0001-6078-786X]{Catherine Walsh}
\affiliation{School of Physics and Astronomy, University of Leeds, Leeds LS2 9JT, UK}

\begin{abstract}
Stable isotopic ratios constitute powerful tools for unraveling the thermal and irradiation history of volatiles. In particular, we can use our knowledge of the isotopic fractionation processes active during the various stages of star, disk and planet formation to infer the origins of different volatiles with measured isotopic patterns in our own solar system. Observations of planet-forming disks with the Atacama Large Millimeter/submillimeter Array (ALMA) now readily detect the heavier isotopologues of C, O and N, while the isotopologue abundances and isotopic fractionation mechanisms of sulfur species are less well understood. Using ALMA observations of the SO and \ce{SO_2} isotopologues in the nearby, molecule-rich disk around the young star Oph-IRS 48 we present the first constraints on the combined \ce{^{32}S}/\ce{^{34}S} and \ce{^{32}S}/\ce{^{33}S} isotope ratios in a planet-forming disk. Given that these isotopologues likely originate in relatively warm gas ($>$50 K), like most other Oph-IRS 48 volatiles, SO is depleted in heavy sulfur while \ce{SO_2} is enriched compared to solar system values. However, we cannot completely rule out a cooler gas reservoir, which would put the SO sulfur ratios more in line with comets and other solar system bodies. We also constrain the \ce{S^{18}O}/SO ratio and find the limit to be consistent with solar system values given a temperature of 60~K. Together these observations show that we should not assume solar isotopic values for disk sulfur reservoirs, but additional observations are needed to determine the chemical origin of the abundant SO in this disk, inform on what isotopic fractionation mechanism(s) are at play, and aid in unravelling the history of the sulfur budget during the different stages of planet formation.
\end{abstract}


\section{Introduction} 

The volatile building blocks of planets, moons and comets exist as gas and ice in protoplanetary disks, which in turn have an origin as ices in cold, dark, molecular clouds \citep[e.g.,][]{2021PhR...893....1O}. On million-year timescales, the material in the disk is subject to a variety of physical and chemical processes that can alter this initially inherited volatile reservoir \citep[e.g.,][]{2022arXiv221214529A}. The isotopic ratios of volatile elements can be used as tools to unravel the history of this material throughout the different stages of the star, disk and planet formation. The main drivers of isotopic fractionation are low temperature chemical reactions, due to the differences in the zero-point vibrational energies of different isotopologues, and isotope selective photo-dissociation (see \citealt{2023ASPC..534.1075N} for a recent review). 

From sub-millimeter observations of planet-forming disks we now have detections of simple molecules containing the heavier isotopes D, \ce{^{13}C}, \ce{^{18}O}, \ce{^{17}O}, \ce{^{15}N} and \ce{^{34}S} \citep[summarised in][]{2023ARA&A..61..287O}. One of the most readily detected isotopes in disks is deuterium and the measured enhancement of D/H in simple molecules (e.g., HCN, \ce{HCO+}, and \ce{N_2H^+}) can be explained via in-situ isotope exchange reactions that are active in the cold outer disk \citep[e.g.,][]{2015ApJ...802L..23F, 2018ApJ...855..119A, 2021ApJS..257...10C, 2023ApJ...943...35M}. The \ce{^{14}N}/\ce{^{15}N} ratio has also been measured in disks for both HCN and CN, where fractionation can likely be attributed to the isotope-selective photo-dissociation of \ce{N_2} in the disk atmosphere \citep{2014A&A...562A..61H, 2017ApJ...836...30G, 2017A&A...603L...6H, 2019A&A...632L..12H, 2018A&A...615A..75V}. Although \ce{C^{18}O} is readily detectable in nearby disks, the optical depth of the line typically prevents a measurement of the oxygen isotope ratio from CO observations alone. Recent work from \citet{2022ApJ...926..148F} shows that optically thin \ce{HCO^+} isotopologue lines can give access to the \ce{^{16}O}/\ce{^{18}O} ratio in disks, which can be used to test mechanisms for recovering the oxygen isotope fractionation observed in the Solar System \citep{2012E&PSL.313...56M}. In the specific case of the TW~Hya disk, the relative abundance of \ce{^{18}O} is found to be consistent with the local interstellar medium value, i.e., is not enhanced or depleted in the disk gas \citep{2022ApJ...926..148F}. 

Overall, the results for Class II disks so far do not require the inheritance of fractionated material from the cold cloud stage. The least understood of the detected volatile elements in disks is sulfur, where both its primary volatile form on the ice and the expected isotopic ratios due to thermal and photo-processing are unclear. The rarer isotopologues of CS, \ce{C^{34}S} and \ce{^{13}CS}, have been detected in a handful of disks, but we lack concrete information on potential in-situ or inherited isotope-fractionation processes that may have occurred \citep{2020ApJ...893..101L, 2019ApJ...876...72L, 2021ApJS..257...12L, 2021A&A...653L...5P, 2023A&A...678A.146B}. However, when considering comets, the remnant icy material from the protoplanetary disk that formed our Solar System, we have significantly more information. The isotopic ratios in sulfur-bearing molecules have been well measured in comet 67P/Churyumov-Gerasimenko (67P/C-G) and there are indicators that isotopic fractionation processes have occurred \citep{2017MNRAS.469S.230P, 2017MNRAS.469S.787C, 2020MNRAS.498.5855A}. In particular, both the \ce{^{32}S}/\ce{^{34}S} and \ce{^{32}S}/\ce{^{33}S} ratios have been derived for a collection of simple molecules and these measurements trace a bulk depletion of heavier isotopes relative to Solar System standards that cannot be explained by photo-dissociation \citep{2017MNRAS.469S.787C}. Additionally, the simple S-bearing molecules SO, \ce{SO_2} and OCS are all enhanced in \ce{^{18}O} by approximately a factor of two relative to solar \citep{2020MNRAS.498.5855A} and \ce{HDS}/\ce{H_2S} is well above the interstellar elemental ratio \citep{2017RSPTA.37560253A}.
The cometary dust particles on the other hand have a \ce{^{32}S}/\ce{^{34}S} ratio consistent with the expected terrestrial value \citep{2017MNRAS.469S.230P}.

The disk with the most observed volatile sulfur, and therefore the source with the most promise for measuring isotopic ratios of sulfur, is the Oph-IRS~48 disk. Oph-IRS~48 is a nearby (135~pc), young Herbig Ae star \citep{2012ApJ...744..116B, Gaia2018} host to a molecule-rich transition disk with an asymmetric dust trap where, both, SO and \ce{SO_2} have been robustly detected \citep{2021A&A...651L...6B, Booth2023_irs48}.  The observable chemistry in this disk appears to be dominated by the sublimation of ices indicated by the presence of \ce{CH_3OH} and other complex organic molecules in the gas phase \citep{2021A&A...651L...5V, 2022A&A...659A..29B, Booth2023_irs48}. This makes the Oph-IRS~48 disk a powerful observational laboratory to investigate the isotopic make-up of the typically hidden icy reservoir in disks. Here, we provide the first measurements of the combined \ce{^{32}S}/\ce{^{34}S} and \ce{^{32}S}/\ce{^{33}S} ratios in a planet-forming disk. We also report a stringent upper-limit on the \ce{S^{18}O} abundance and use this to estimate the corresponding lower-limit of the \ce{^{16}O}/\ce{^{18}O} isotopic ratio. 

\begin{figure*}[t!]
    \centering
    \includegraphics[width=\hsize]{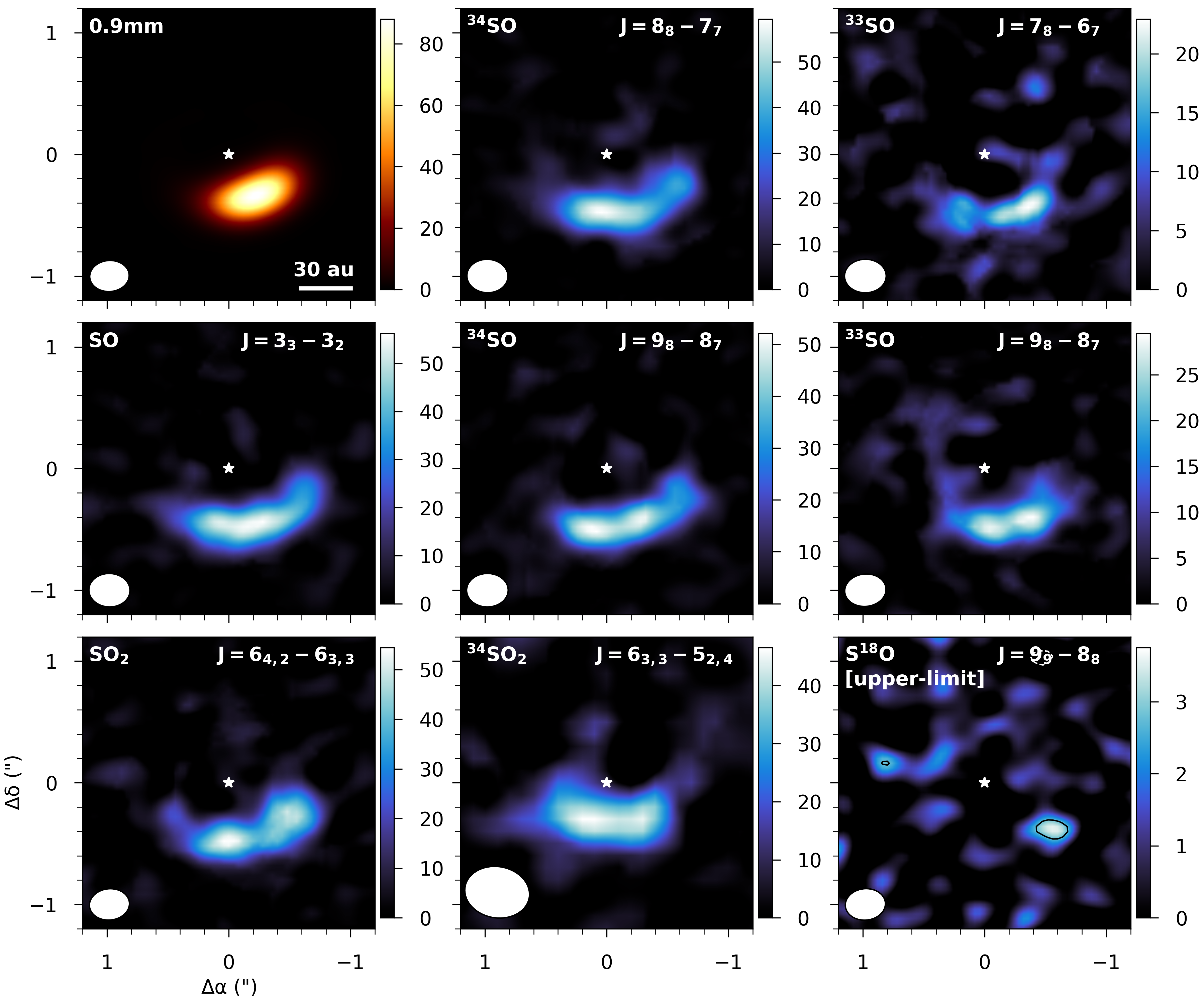}
    \caption{Integrated intensity maps of the 0.9~mm dust continuum emission and the sulfur isotopologues in the Oph-IRS~48 disk. The units of the color bar are mJy~beam$^{-1}$~km~$\mathrm{s^{-1}}$ for the molecular lines and mJy~beam$^{-1}$ for the continuum. The black contours on the \ce{S^{18}O} map highlight 3$\times$ the rms noise level.}
    \label{fig:fig1}
\end{figure*}

\section{Methods}

We use molecular line observations of the Oph-IRS~48 planet-forming disk taken with the Atacama Large Millimeter/sub-millimeter Array (ALMA) from the programs 2017.1.00834.S (PI: Adriana Pohl) and 2021.1.00738.S (PI: Alice S. Booth). The former were first presented in \citet{2021A&A...651L...5V} and \citet{2021A&A...651L...6B} and the latter in \citet{Booth2023_irs48}.  The 2017.1.00834.S data were pipeline calibrated by \citet{2020ApJ...900...81O} and \citet{2021A&A...651L...6B} imaged these data with CASA task \texttt{tCLEAN} \citep{2007ASPC..376..127M} at a velocity resolution of 1.7~km~s${^{-1}}$ and a Briggs robust weighting of 0.5 resulting in a $\approx$0.45$\farcs$ beam. The 2021.1.00738.S data were self-calibrated (both phase and amplitude) and imaged with CASA task \texttt{tCLEAN} at a velocity resolution of 0.9~km~s${^{-1}}$ with a Briggs robust weighting of 0.5 resulting in a $\approx$0.3$\farcs$ beam \citep{2021A&A...651L...5V}. All integrated intensity maps were generated using Keplerian masks \citep{rich_teague_2020_4321137}. In the above referenced works, the data calibration, line identification and imaging steps are described in full detail.

The focus of this paper is on the sulfur-bearing molecules covered in the data. \citet{2021A&A...651L...6B} and \citet{Booth2023_irs48} report the detections of SO, \ce{^{34}SO}, \ce{^{33}SO}, \ce{SO_2} and \ce{^{34}SO_2} and we use the integrated intensity maps from these works.  One transition of \ce{S^{18}O} with a $\mathrm{A_{ul}}\approx10^{-3.2}$~s$^{-1}$ is covered in the data from \citet{Booth2023_irs48} and we tentatively detect this line in one velocity channel (0.9~km~$\mathrm{s^{-1}}$) at a signal-to-noise ratio of 4. Although only detected in one channel, this is the same velocity channel where the main SO isotope lines are the brightest and the local line-width per pixel is only $\approx$2-3 km~s$^{-1}$. 
A robust detection would require 3$\sigma$ emission over $>$3 channels to confirm. We therefore consider this as an upper-limit on the \ce{S^{18}O} abundance in this disk. The channel maps of the \ce{S^{18}O} line are shown in Figure~\ref{fig:channels}.
This \ce{S^{18}O} line was imaged in the same manner as the other lines from \citet{Booth2023_irs48} and is part of the same ALMA data set. Table~\ref{tab:tab1} lists the properties of the transitions and images presented in this paper.

Under the typical assumptions of optically thin line emission and local thermodynamic equilibrium \citep[as done by][]{2018ApJ...859..131L}, we estimate the peak column density traced by the SO and \ce{SO_2} isotopologues detected in the Oph-IRS~48 disk. The various line properties and partition functions of each isotopologue are taken from the Cologne Database for Molecular Spectroscopy \citep[CDMS;][]{2001A&A...370L..49M, 2005JMoSt.742..215M, 2016JMoSp.327...95E} with data from \citet{1985JPCRD..14..395L, 1992ApJ...399..325L, 1996JMoSp.180..197K}. The errors in the peak column densities are propagated from the errors in the integrated intensity maps (as listed in Table~\ref{tab:tab1}) where these errors are a measure of the root mean square noise (rms) of the line-free regions of the images without Keplerian masking. The rms in the \ce{S^{18}O} is lower than the other lines since this integrated map is only calculated for one channel. 

We calculate these column densities at a range of excitation temperatures (30 to 150~K in steps of 10~K) which cover the range of expected excitation conditions in this disk. The lower limit of this range is informed by the brightness temperature of the midplane millimeter dust which is at a minimum of 30~K and the upper limit is from the $\gtrsim$100~K brightness temperature of the optically thick CO isotopologues, which trace the disk surface layers \citep{2021A&A...651L...5V, 2023A&A...673A...7L} and, the rotational temperatures of \ce{H_2CO}, \ce{CH_3OH} and \ce{SO_2} (\citealt{2021A&A...651L...5V} and Temmink et al. in prep.). In particular, using the line survey data of the Oph-IRS~48 disk presented in \citet{Booth2023_irs48}, where $>$10 \ce{SO_2} lines are detected with an upper energy level range of 35 to 200~K, Temmink et al. 2024 (in prep.) find that the rotational temperature for \ce{SO_2} to be very well constrained to $\approx$60~K at the emission peak. As the SO and \ce{SO_2} emission is co-spatial in the observations (as shown in Figure~\ref{fig:fig1}) and chemically linked in disk models through their origins in the warm molecular layer, not the disk midplane \citep[e.g.,][]{2024MNRAS.528..388K}, we expect them to trace the same gas temperature.
Additionally, to support this assumption of warm gas, in disk models, SO and \ce{SO_2} freeze out at dust temperatures less than $\approx$40~K and $\approx$60~K, respectively \citep[e.g.,][]{2018A&A...617A..28S}. We note that this will depend on the assumed binding energies for example, \citet{2021arXiv211006010P} used quantum chemical calculations to show that SO has a range of binding energies on a water ice cluster (60 molecules), spanning a temperature range of 933 to 3483 K with an average value of 2128 K, which is lower than the values listed in UDfA and KIDA (2800 and 2600 K, respectively). 
Ideally, we would have an independent measure of the SO rotational temperature but in total, \citet{Booth2023_irs48} detect only three SO lines in the Oph-IRS~48 disk and the high line opacities of the $J=7_8-6_7$ and $J=8_8-7_7$ transitions do not allow for a rotation diagram analysis. In this work, we only use the optically thinnest transition detected ($J=3_3-3_2$) as a measure of the SO column density. For \ce{^{34}SO} and \ce{^{33}SO} two lines of each are detected and we take the average column density traced by these lines at each temperature. Therefore, although we expect the SO to trace warm 60~K gas, we cannot make an independent temperature measurement to confirm this, therefore, we explore a range of higher and lower gas temperatures to see how sensitive the inferred isotopic ratios is to this assumption. Due to the different intrinsic properties of the transitions of each isotopologue these ratios will vary with the assumed gas temperature.

% , the rotational temperatures of other molecules and the inferred disk temperature structure \citep{2021A&A...651L...5V, 2023A&A...670A.154W, 2023A&A...673A...7L}. 

To compare our isotope ratios with measurements of isotopic variations in Solar System bodies, we also calculate the deviation of our ratios relative to the Vienna Canyon Diablo Troilite (V-CDT). This is the most common isotopic standard used in comet and meteorite studies. The standard ratios used are 22.64 for \ce{^{32}S}/\ce{^{34}S}, 126.9 for \ce{^{32}S}/\ce{^{33}S} and 498.7 for \ce{^{16}O}/\ce{^{18}O} \citep{2001GeCoA..65.2433D, 2001RCMS...15..501W}. These differ slightly from the local ISM ratios, where in particular, \ce{^{16}O}/\ce{^{18}O} is 550 \citep{1994ARA&A..32..191W}. 
The deviations ($\delta$) are reported in units of per mille (\textperthousand) and are calculated as followed:
$$\delta\ce{^{34}S} = [(\ce{^{34}S}/\ce{^{32}S})_{disk}/(\ce{^{34}S}/\ce{^{32}S})_{standard} -1]~\times~1000$$
where the above is an example for the \ce{^{34}S}/\ce{^{32}S} ratio. 

\section{Results} 

In Figure~\ref{fig:fig1}, we present the integrated intensity maps of sulfur isotopologues in the Oph~IRS~48 disk and the 0.9~mm continuum emission. The molecular species all appear co-spatial and are only present in the Southern region of the disk, where the millimeter dust grains are most prominent. From these integrated intensity maps, we take the peak integrated flux and use this to calculate the molecular column density traced by each of the isotopologues. The resulting peak column densities, corresponding isotopic ratios, and deviations from V-CDT at 30, 60 and 100~K are listed in Table~\ref{tab:tab2} and across the full temperature range in Figure~\ref{fig:fig2}. We also compared these peak column densities to the averaged column densities derived from the whole SO and \ce{SO_2} emitting region and find that the resulting isotope ratios, the key result of this paper, are within the quoted 1~$\sigma$ errors.

In Figure~\ref{fig:fig4}, we show the inferred ratios as a function of the assumed temperature. The isotope ratios derived from \ce{^{32}SO} all increase with increasing gas temperature and both \ce{^{34}S} and \ce{^{33}S} are consistent with the V-CDT values at temperatures lower than $\approx$50~K. When adopting our fiducial temperature of 60~K, we find that traced in SO line emission, the Oph-IRS~48 disk is depleted in both \ce{^{34}S} and \ce{^{33}S} with \ce{^{32}S}/\ce{^{34}S}=$34\pm4$ and \ce{^{32}S}/\ce{^{33}S}=$290\pm60$. The \ce{^{16}O}/\ce{^{18}O} lower limit from SO of $>$500 is consistent with the V-CDT value. The SO $J=3_3-3_2$ transition is optically thin with $\tau \approx 0.1$ and since the rare isotopes of SO are depleted relative to this, optical depth cannot explain the anomalous 
%\movetabledown=2cm
\begin{rotatetable*}
    \begin{deluxetable*}{c c c c c c c c c c}
    \tablecaption{Molecular line and image properties where molecular transition data are taken from CDMS \citep{2005JMoSt.742..215M, 2016JMoSp.327...95E}.}
    \tablehead{\colhead{Mol.} & \colhead{Transition} & \colhead{Frequency} & \colhead{E$\mathrm{_{up}}$} & \colhead{log$_{10}$(A$_{\mathrm{ul}}$)} & \colhead{Beam (Position Angle)} & \colhead{Peak Integrated Intensity} & \colhead{Uncertainty} \\
&& \colhead{[GHz]} & \colhead{[K]} & \colhead{[s$^{-1}$]} &   \colhead{[\farcs$\times$\farcs ($^{\circ}$)]} &   \colhead{[mJy/beam km/s]}  &   \colhead{[mJy/beam km/s]}}
    \startdata
    \ce{SO} &  $J=3_{3}-3_{2}$         & 339.3414590 & 25.5 & -4.8372          & 0.34$\times$0.28 (88)  & 56.2 & 4.7 \\
    \ce{^{34}SO} &  $J=8_{8}-7_{7}$         & 339.8572694 & 77.3 & -3.2944     & 0.34$\times$0.27 (88)  & 58.7 & 4.7 \\
    \ce{^{34}SO} &  $J=9_{8}-8_{7}$         & 337.5801467 & 86.1 & -3.3109     & 0.34$\times$0.28 (-89) & 52.1 & 5.2 \\
    \ce{^{33}SO}$^{*}$ &  $J=7_{8}-6_{7}$   & 337.1986199 & 80.5 & -3.3158     & 0.34$\times$0.28 (-90) & 22.9 & 5.3 \\
    \ce{^{33}SO}$^{*}$ &  $J=9_{8}-8_{7}$   & 343.0882949 & 78.0 & -3.2819     & 0.34$\times$0.27 (-85) & 29.5 & 4.7 \\
    \ce{S^{18}O} & $J=8_{9}-7_{8}$          & 355.5735893 & 93.1 & -3.2414    & 0.33$\times$0.26 (-85) & 3.6  & 0.9 \\
    \ce{SO_2} & $J=6_{4,2}-6_{3,3}$         & 357.9258478 & 58.6  & -3.5845    & 0.33$\times$0.26 (-85) & 50.1 & 5.9 \\
    \ce{^{34}SO_2} & $J=6_{3,3}-5_{2,4}$    & 362.1582327 & 40.7 & -3.4839     & 0.53$\times$0.42 (80)  & 46.5  & 5.0  \\
    \enddata
    \tablecomments{$^{*}$The \ce{^{33}SO} lines are blends of multiple hyper-fine components. Here, we show the properties of the strongest transition and the flux is the total of the blended lines. The $J=7_{8}-6_{7}$ line includes the F=17/2-15/2, F=15/2-13/2, F=11/2-9/2 and F=11/2-9/2 hyper-fine components and the $J=9_{8}-8_{7}$ line includes the F=21/2-19/2, F=19/2-17/2, F=17/2-15/2 and F=15/2-13/2 hyper-fine components.}
    \label{tab:tab1}
    \end{deluxetable*}
\end{rotatetable*} 
\clearpage 
\noindent isotope ratios we recover when assuming our fiducial temperature of 60~K or a temperature above this. Additionally, the \ce{^{34}SO}/\ce{^{33}SO} ratio is approximately constant with temperature and is always 40-50\% above the V-CDT value.

The \ce{SO_2} isotope ratios varies less with temperature, since the upper energy levels of the \ce{SO_2} and \ce{^{34}SO_2} lines are similar. In contrast to our findings for SO, the \ce{^{34}S} column density appears to be enhanced at all temperatures when measured from the \ce{SO_2} and at 60~K the \ce{^{32}S}/\ce{^{34}S}=5. From a rotation diagram analysis, Temmink et al. in prep. show that \ce{^{32}SO_2} emission is optically thin, but, because the observations are spatially and kinematically unresolved, the optical depth is likely underestimated. Therefore, the inferred \ce{^{32}S}/\ce{^{34}S} ratio from \ce{SO_2} may only be a lower limit.

\section{Discussion} 

The currently observed volatile sulfur reservoir in the Oph-IRS~48 disk is primarily in the form of SO. This is different to most Class II disks where generally CS is the most common S-species detected \citep[e.g.,][]{2019ApJ...876...72L, 2021ApJS..257...12L}. CS is detected in the Oph-IRS~48 disk, but its column density is $\approx$1000$\times$ lower than for SO; therefore, CS makes a negligible contribution to the total S abundance \citep{Booth2023_irs48}. In Figure~\ref{fig:fig3}, we show the contributions from \ce{H_2S}, S, OCS, SO and \ce{SO_2} to the currently total observed volatile sulfur reservoirs in the low-mass embedded Class 0 protostar IRAS 16293-2442~B, the more evolved Class II disk Oph-IRS~48, the comet 67P/C-G and the shocked region L1157-B1 \citep[with values from][]{2016MNRAS.462S.253C, 2018MNRAS.476.4949D, 2019ApJ...878...64H, 2023A&A...672A.122K, Booth2023_irs48}. 

The differences in sulfur reservoirs between IRAS~16293-2442~B and 67P/C-G are discussed in detail in \citet{2018MNRAS.476.4949D}, who suggest that UV-irradiation during the star formation process converted \ce{H_2S} to OCS. OCS is not detected in Oph-IRS~48 down to $<$1\% of the SO column density and this is unexpected, as discussed in \citet{Booth2023_irs48}, due to the other signatures of ice sublimation in this disk. Interestingly, the \ce{SO_2}/OCS ratio we see in Oph-IRS~48 is also significantly higher than that measured for the outbursting source V883~Ori, where ice sublimation is also occurring \citep{2024AJ....167...66Y}. Oph-IRS~48 is a young Herbig Ae star and therefore the volatiles, in the gas and ice phases, will be subjected to UV-irradiation, which could chemically transform the primary S-reservoir in the disk. It is also possible that in the Oph-IRS~48 disk there are shocks liberating a more refractory form of sulfur, e.g., \ce{S_2}, that when released from the grains goes on to form SO (and \ce{SO_2}) efficiently in the gas phase (as discussed in \citealt{2024MNRAS.528..388K}). Indeed, the bulk of the S in disks is expected to be in refractory form \citep{2019ApJ...885..114K}.  The high SO/\ce{SO_2} ratio in Oph-IRS~48 is also most closely a match with that in the shocked region L1157-B1, but the CS/SO ratio is orders of magnitude higher in the shocked region when compared to the Oph-IRS~48 disk \citep{2019ApJ...878...64H}. Observations of the shock tracer SiO (and SiS, \citealt{2023ApJ...952L..19L}) and key sulfur species \ce{H_2S} will help unravel the distinctions between the volatile reservoir in the Oph-IRS~48 disk and that in other environments.

\begin{table*}[t!]
\centering
\caption{Derived peak column densities and isotope ratios with associated 1~$\sigma$ errors. Note that 60~K is our fiducial temperature. The V-CDT standard values are from \citet{2001GeCoA..65.2433D} and \citet{2001RCMS...15..501W}.}
\begin{tabular}{ccccc}
\hline \hline
Molecule       &     & $\mathrm{N_{peak}}$  [$\mathrm{cm^{-2}}$]  &      \\ 
                & $\mathrm{T_{ex}}$: 30~K          & \textbf{60~K} & 100~K \\ \hline
\ce{^{32}SO}    &  2.7$\pm0.2\times10^{15}$ &   3.6$\pm0.3\times10^{15}$   &   5.1$\pm0.41\times10^{15}$\\
\ce{^{34}SO}    &  2.0$\pm0.2\times10^{14}$ &   1.0$\pm0.1\times10^{14}$   &   1.0$\pm0.09\times10^{14}$\\
\ce{^{33}SO}    &  2.2$\pm0.4\times10^{13}$ &   1.2$\pm0.2\times10^{13}$   &   1.3$\pm0.25\times10^{13}$\\
\ce{S^{18}O}    &  1.5$\pm0.4\times10^{13}$ &   0.7$\pm0.2\times10^{13}$   &   0.7$\pm0.17\times10^{13}$\\
\ce{SO_2}       &  6.9$\pm0.4\times10^{14}$ &   6.7$\pm0.5\times10^{14}$   &   9.6$\pm0.69\times10^{14}$\\
\ce{^{34}SO_2}  &  0.9$\pm0.1\times10^{14}$ &   1.3$\pm0.1\times10^{14}$   &   2.1$\pm0.21\times10^{14}$\\ \hline
Ratio       & $\mathrm{T_{ex}}$: 30 K &textbf{60~K} & 100~K &    V-CDT \\ \hline
\ce{^{32}SO}/\ce{^{34}SO}       & $13\pm1$ &   $34\pm4.0$     & $50\pm6.0$      & 22.64 \\ 
\ce{^{32}SO_2}/\ce{^{34}SO_2}   & $7.5\pm0.9$  &    $5.2\pm0.6$     & $4.6\pm0.6$       & 22.64 \\ 
    \ce{^{32}SO}/\ce{^{33}SO}   & $120\pm25$ &  $290\pm60$  & $410\pm90$                & 126.9 \\
  \ce{^{32}SO}/\ce{S^{18}O}       & $>180$ &  $>$500 & $>$760   & 498.7 \\ 
% \ce{^{32}SO}/\ce{S^{18}O}       & $180\pm50$ &  $500\pm130$ & $760\pm200$               & 498.7 \\ 
\hline 
Deviation from V-CDT       & $\mathrm{T_{ex}}$: 30 K & textbf{60~K} & 100~K &    \\ \hline
$\delta$\ce{^{34}SO}       &   690$^{+160}_{-200}$ &  -340$^{+90}_{-70}$      &  -550$^{+60}_{-50}$     &     \\ 
$\delta$\ce{^{34}SO_2}     &  2016$^{+410}_{-320}$ &  3350$^{+250}_{-170}$    &  3920$^{+740}_{-570}$     &     \\ 
$\delta$\ce{^{33}SO}       &  23$^{+250}_{-170}$          & -559$^{+120}_{-80}$       &  -690$^{+80}_{-60}$    &     \\ 
% $\delta$\ce{S^{18}O}        &   $<$1820$^{-570}_{-960}$         &  $<$5$^{+360}_{-210}$      &  $<$-340$^{+230}_{-140}$    &     \\ 
$\delta$\ce{S^{18}O}        &   $<$1820        &  $<$5      &  $<$-340    &     \\ 
\hline 
\end{tabular}
\label{tab:tab2}
\end{table*}

\begin{figure*}
    \centering
    \includegraphics[width=\hsize]{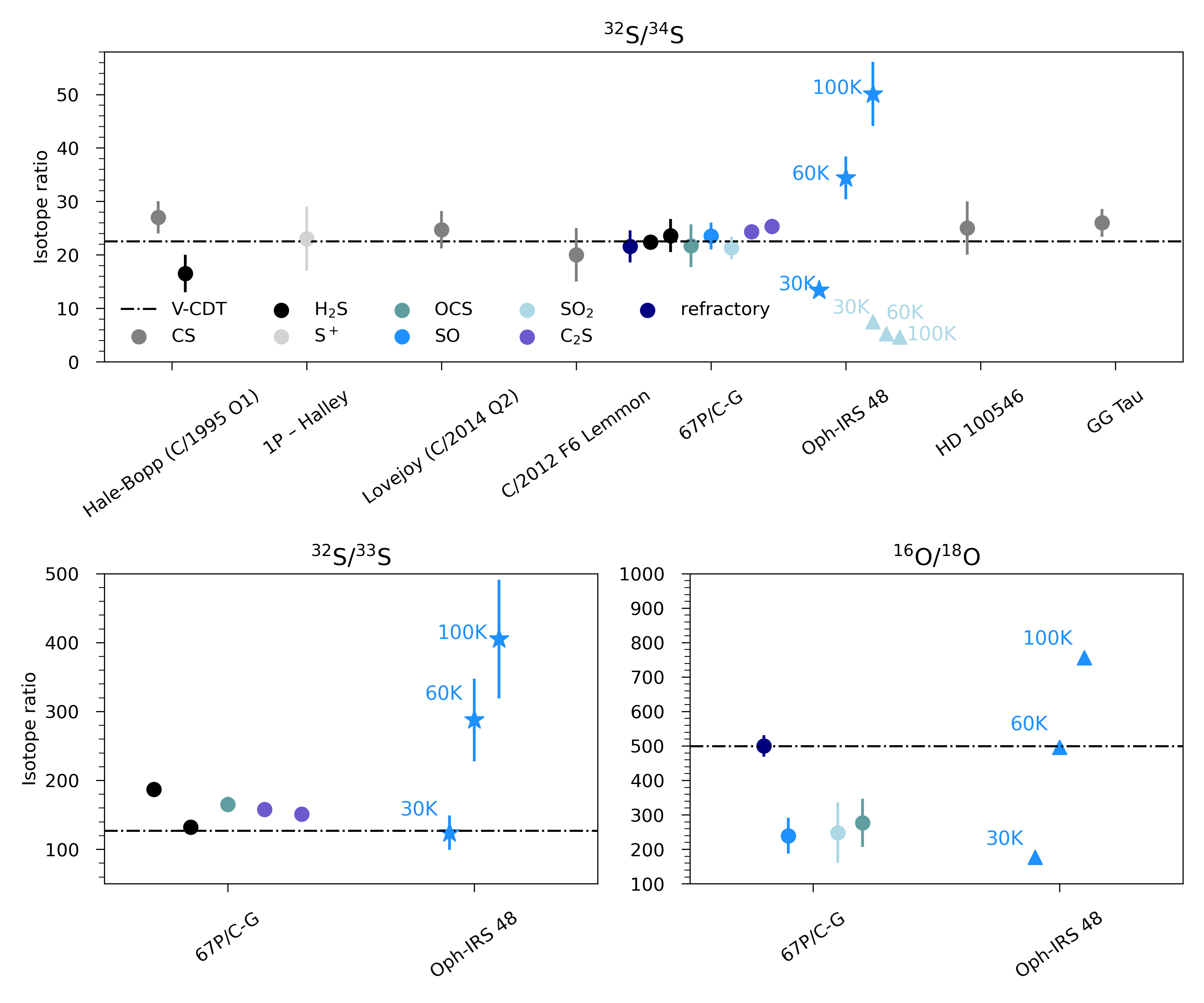}
    \caption{Inferred isotopic ratios of S and O in different volatile sulfur species detected towards comets and protoplanetary disks. The results for Oph-IRS~48 are highlighted by the star or arrow markers at temperatures of 30, 60 and 100~K. 
    Note the \ce{^{34}S} ratio from \ce{SO_2} and the \ce{^{18}O} ratio from SO are lower limits.
    The dashed lines denote the V-CDT standard isotopic ratios for \ce{^{32}S}/\ce{^{34}S}, \ce{^{32}S}/\ce{^{33}S} and \ce{^{16}O}/\ce{^{18}O} \citep{2001GeCoA..65.2433D, 2001RCMS...15..501W}. The vertical bars mark the $\pm$1~$\sigma$ errors.
    The other disk values are taken from \citet{Booth2023_hd100546,2019ApJ...876...72L} and \citet{2021A&A...653L...5P}.
    Cometary values are taken from \citet{1997Sci...278...90J, 2004A&A...418.1141C, 2016A&A...589A..78B, 2017MNRAS.469S.787C, 2017MNRAS.469S.230P, 2018MNRAS.477.3836P} and \citet{2020MNRAS.498.5855A}.
    We exclude the \ce{S_2} isotope ratio reported in \citet{2015PhDT.......418C} and subsequently in \citet{2017MNRAS.469S.230P} for 67P/C-G from our comparison due to 
    the low signal-to-noise of the \ce{S_2} peak in the mass spectra and the significant contamination from other overlapping species (priv. comm. Martin Rubin).}
    \label{fig:fig2}
\end{figure*}

\begin{figure*}[t!]
    \centering
    \includegraphics[width=0.95\hsize]{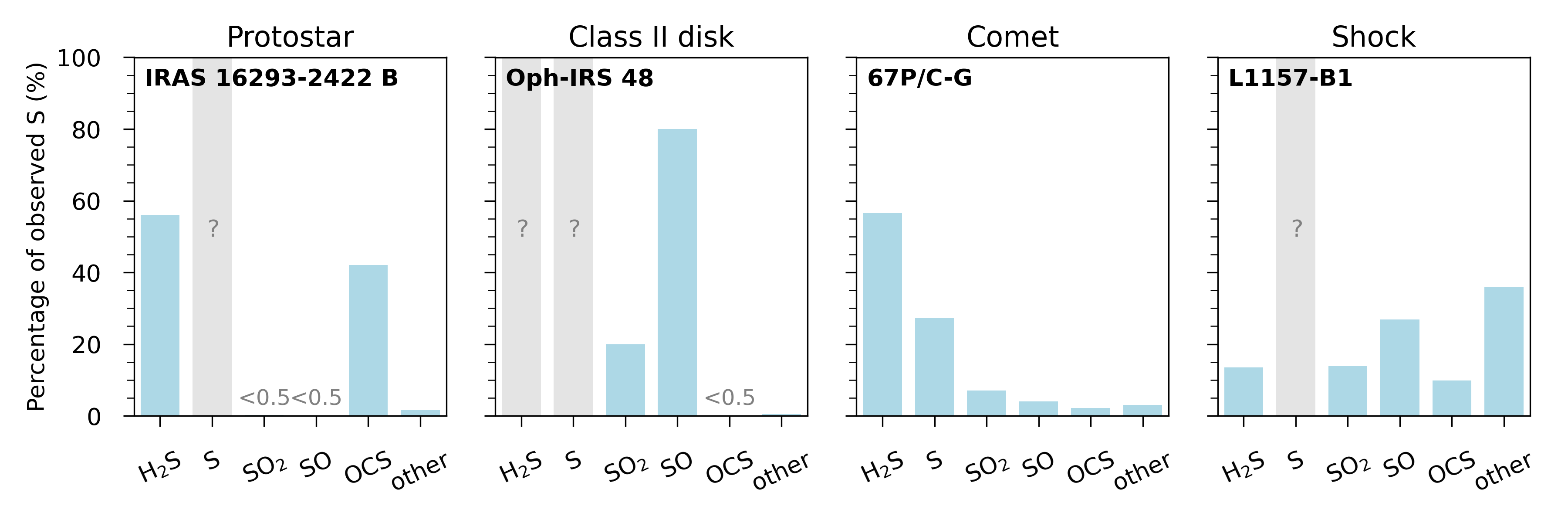}
    \caption{Bar plots showing the percentage of observed gas-phase sulfur in different atoms/molecules for the protostar IRAS 16293-2442~B, Class II disk Oph-IRS~48, comet 67P/C-G and shocked region L1157-B1. Values are taken from \citet{2016MNRAS.462S.253C, 2018MNRAS.476.4949D, 2019ApJ...878...64H, 2023A&A...672A.122K} and \citet{Booth2023_irs48}. Where the percentage is less than 0.5\%, a note is added. We note that this comparison naively assumes that 100\% of the volatile S can be accounted for in the currently detected molecules in these systems. }
    \label{fig:fig3}
\end{figure*}

Given the differences in molecular reservoirs of S we may expect differences in the isotopic ratios on volatile sulfur in the Oph-IRS48 disk. To place our isotopic ratios in context, in Figure~\ref{fig:fig2}, we show a comparison of the isotope ratios measured in comets to Class II disks for a variety of simple species. 
The ratios we measure do vary as a function of the assumed gas temperature (as shown in Figures~\ref{fig:fig2} and \ref{fig:fig4}). If the gas is cool ($<$50~K), the \ce{^{34}S} and \ce{^{33}S} ratios for SO are within range 1-2$\sigma$ range of the V-CDT values. This would mean there are no significant isotopic differences between the Oph-IRS~48 disk and the solar system and, that volatile S is not severely fractionated during the planet formation timescale. Similarly, at 30~K we would tentatively recover the \ce{^{18}O} enhancement as seen in 67P/C-G but further data are needed to verify this.

The rotational temperature of \ce{SO_2} in this disk has been well constrained to be 60~K (Temmink et al. in prep.) therefore, we discuss the case where the SO gas is also warm, as also supported by astrochemical models \citep{2018A&A...617A..28S,2024MNRAS.528..388K}.
At temperatures of 60~K and above the isotope ratios for sulfur are anomalous when compared to the V-CDT and cometary values. This possible observed depletion of \ce{^{34}S} and \ce{^{33}S} traced in SO is currently not traced in any cometary volatiles and in particular, not in the \ce{^{34}SO} measured in 67P/C-G \citep{2020MNRAS.498.5855A}. This depletion is also not traced in the refractory dust of 67P/C-G nor the dust collected in the STARDUST mission \citep{2012M&PS...47..649H, 2017MNRAS.469S.230P}. 
For the other few disks with robust measurements of \ce{^{32}S}/\ce{^{34}S} from CS, these values are consistent with the expected isotope ratio of 22 \citep{2021A&A...653L...5P,Booth2023_hd100546}.
In the broader context of meteorite samples, our $\delta^{34}$S and $\delta^{33}$S values, calculated at 60~K, lie at the most depleted end of the distribution of isotopic ratios measured in SiC grains from the meteorites Murchison and Indarch \citep[this literature is summarised in][]{2017MNRAS.469S.787C}. The apparent factor $\approx$4 enhancement of \ce{^{34}S} in \ce{SO_2} is an extreme outlier in all cases, but this may be due to beam diluted and optically thick \ce{^{32}SO_2} emission. 
This low \ce{^{34}S} ratio in \ce{SO_2} is similar to that measured in SO for Class 0/I sources in Perseus \citep{2023A&A...678A.124A} where this can be attributed to optically thick emission from the main isotopologue. 

Finally, when considering the lower limit of the \ce{^{18}O} isotope ratio in SO at $\gtrsim$60~K find a difference to the volatiles in 67P/C-G. The enhanced \ce{^{18}O} in SO, \ce{SO_2} and OCS in 67P/C-G and, the high HDS/\ce{H_2S} (of $\approx10^{-3}$) indicate the possible pre-solar formation of these molecules in the cold dark cloud stage \citep{2017RSPTA.37560253A}. In comparison, for warm gas, in the Oph-IRS~48 disk our upper-limit does not indicate any oxygen isotope fractionation, similar to the findings from \citet{2022ApJ...926..148F} for \ce{H^{13}CO^+}/\ce{HC^{18}O^+} in TW~Hya. 
The nominal ratio of $\approx$500 is also the same as that measured for the 67P/C-G cometary dust particles \citep{2018MNRAS.477.3836P}.

The observations of Oph-IRS~48 show the possibility of isotopically distinct sulfur reservoirs in planet forming disks with an isotopic fractionation distinct to that of 67P/C-G. 
The depletion of heavier isotopes and lack of \ce{^{18}O} enhancement would also suggest that low temperature isotope exchange reactions are not responsible as these would have the opposite effect. If the optically thin nature of \ce{SO_2} holds, the different isotopic signatures in SO and \ce{SO_2} mean that they may not share a common formation path via \ce{H_2S} and \ce{H_2O} ice sublimation \citep{2018A&A...617A..28S, 2021A&A...651L...6B}. 
If SO and \ce{SO_2} both form from photodissociated \ce{H_2S}, they would be expected to have the same isotopic ratios. The isotope selective photo-dissociation of \ce{H_2S} has been proposed as a mechanism for the enrichment of heavy isotopes in some chondritic meteorites, but if this were the case in Oph-IRS~48, we would expect both the SO and \ce{SO_2} to be enriched in the heavier isotopes, if they indeed formed from \ce{H_2S} \citep{2013PNAS..11017650C, 2017GeCoA.196..326L, 2021GeCoA.309..135V}. If there are shocks in the Oph-IRS~48 disk, the injection of refractory S into the gas phase could pollute the isotopic ratios set by photo-processes. Future observations of Oph-IRS~48 to detect \ce{H_2S} and measure its isotopic ratios will be necessary to determine if SO and/or \ce{SO_2} are daughter molecules forming from \ce{H_2S}. Additionally, observations of multiple optically thin lines of SO and its isotopologues will enable an independent temperature measurement for the SO rather than assuming the same temperature as \ce{SO_2}. 

\section{Conclusion} 

We have presented the first measurements of the combined \ce{^{34}S} and \ce{^{33}S} isotope ratios relative to \ce{^{32}S} in a planet-forming disk. Additionally, we share the an upper-limit on the \ce{S^{18}O} column density and the inferred lower-limit on the \ce{^{16}O}/\ce{^{18}O} ratio. The observed isotopic fractionation of the S reservoir in the Oph-IRS~48 disk is potentially distinct from that measured in comets for both the elements sulfur and oxygen but, this relies heavily on the adopted gas temperature. Assuming a kinetic temperature of 60~K (as measured for \ce{SO_2}) or above, both \ce{^{34}S} and \ce{^{33}S} are depleted relative to their Solar System standard values when traced in SO. At cooler temperatures, the isotope ratios are consistent with the cometary values indicating the need for additional data to better and independently constrain the SO gas temperature.
Interestingly, at all temperatures the \ce{^{34}S}/\ce{^{33}S} ratio traced in SO is enhanced.
We also identify a tentative enhancement in the heavier \ce{^{34}SO_2} isotopologue which requires observations of optically thinner \ce{^{32}SO_2} transitions to confirm. 

There are also possible limitations in the current data due to dust optical depth that will require longer wavelength observations to assess. It may be that the isotopologues are tracing gas deeper in the disk, below the optically thick dust surface, resulting in an underestimated column density measurement, as seen towards protostars \citep{2020ApJ...896L...3D}. Furthermore, follow-up observations with ALMA to detect \ce{H_2S} (and its isotopologues) will help to complete the observable volatile inventory of S in this disk. Knowing if \ce{H_2S} is depleted or enhanced in the heavier isotopes will aid significantly in understanding the isotopic fractionation mechanisms at play. More generally, future high-sensitivity line surveys of populations of protoplanetary disks will be able to target multiple isotopologues of S and other volatile elements, which can be used to connect the dots between the initial and final stages of planetary system formation. 

\clearpage
\begin{acknowledgements}
This paper makes use of the following ALMA data: 2017.1.00834.S, 2021.1.00738.S. We acknowledge assistance from Allegro, the European ALMA Regional Centre node in the Netherlands. ALMA is a partnership of ESO (representing its member states), NSF (USA) and NINS (Japan), together with NRC (Canada), MOST and ASIAA (Taiwan), and KASI (Republic of Korea), in cooperation with the Republic of Chile. The Joint ALMA Observatory is operated by ESO, AUI/NRAO and NAOJ. 
This work has used the following additional software packages that have not been referred to in the main text: Astropy, IPython, Jupyter, Matplotlib and NumPy \citep{Astropy,IPython,Jupyter,Matplotlib,NumPy}.
Astrochemistry in Leiden is supported by funding from the European Research Council (ERC) under the European Union’s Horizon 2020 research and innovation programme (grant agreement No. 101019751 MOLDISK).
A.S.B. is supported by a Clay Postdoctoral Fellowship from the Smithsonian Astrophysical Observatory. 
M.N.D. acknowledges the Holcim Foundation Stipend. 
M.T. acknowledges support from the ERC grant 101019751 MOLDISK. 
Support for C.J.L. was provided by NASA through the NASA Hubble Fellowship grant No. HST-HF2-51535.001-A awarded by the Space Telescope Science Institute, which is operated by the Association of Universities for Research in Astronomy, Inc., for NASA, under contract NAS5-26555.
M.L. is funded by the European Union (ERC, UNVEIL, 101076613). Views and opinions expressed are however those of the author(s) only and do not necessarily reflect those of the European Union or the European Research Council. Neither the European Union nor the granting authority can be held responsible for them.
S.N.~is grateful for support from RIKEN Special Postdoctoral Researcher Program (Fellowships), Grants-in-Aid for JSPS (Japan Society for the Promotion of Science) Fellows Grant Number JP23KJ0329, and MEXT/JSPS Grants-in-Aid for Scientific Research (KAKENHI) Grant Numbers JP 18H05441, JP20K22376, 
JP20H05845, JP20H05847, JP23K13155, and JP24K00674. 
C.W.~acknowledges financial support from the University of Leeds, the Science and Technology Facilities Council, and UK Research and Innovation (grant numbers ST/X001016/1 and MR/T040726/1).
\end{acknowledgements}

\bibliography{sample631}{}
\bibliographystyle{aasjournal}

\newpage
\appendix
\section{\ce{S^{18}O} channel maps}
\begin{figure*}[h!]    
    \centering
    \includegraphics[width=\hsize]{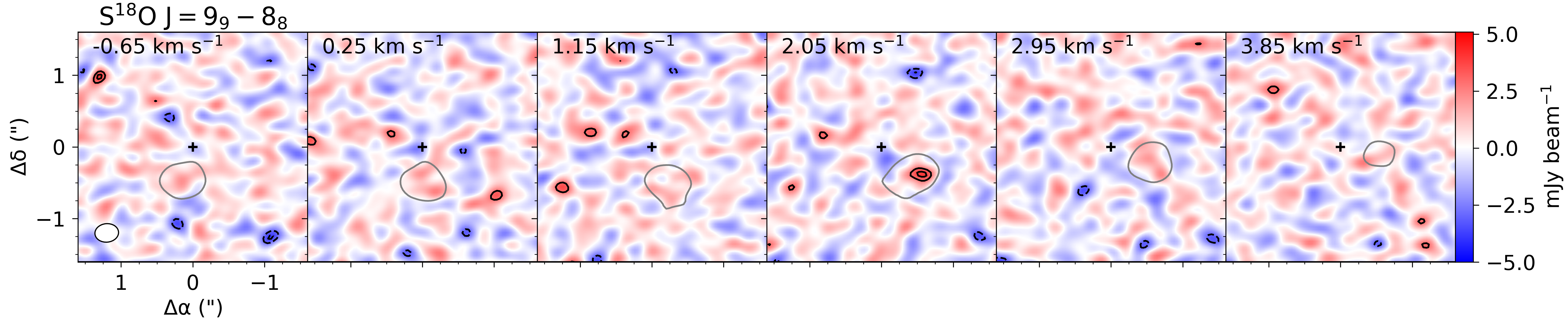}
    \caption{Channel maps of the \ce{S^{18}O} $J=9_9-8_8$ transition where the black solid contours mark the [3, 4]$\times~\sigma$ level and black dashed contours mark [-4, -3]$\times~\sigma$ level. The grey solid contour marks the 5$\sigma$ level of the \ce{^{34}SO} $J=9_8-8_7$ transition to show the expected position velocity pattern of the disk emission.}
    \label{fig:channels}
\end{figure*}

The channel maps generated over the frequency range of the \ce{S^{18}O} $J=9_9-8_8$ transition are shown in Figure~\ref{fig:channels}.

\newpage
\section{Isotope ratios as a function of temperature}
\begin{figure*}[h!]
    \centering
    \includegraphics[width=0.9\hsize]{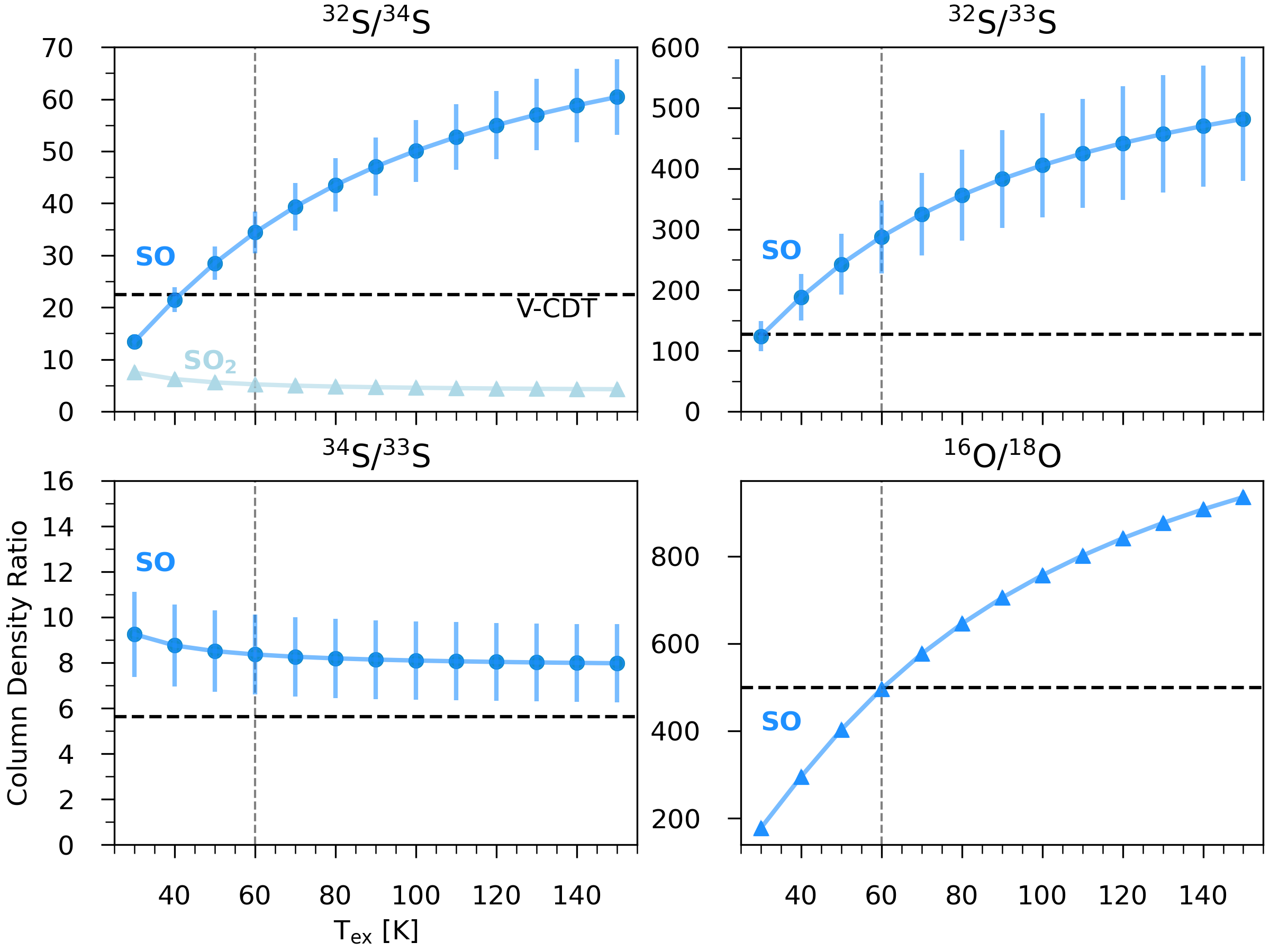}
    \caption{Inferred isotopic ratios of S and O in SO and \ce{SO_2} for the Oph-IRS~48 disk calculated for a range of excitation temperatures.
    Note the \ce{^{34}S} ratio from \ce{SO_2} and the \ce{^{18}O} ratio from SO are lower limits.
    The vertical dashed lines highlight our fiducial temperature of 60~K, which is the rotational temperature of \ce{SO_2} (Temmink et al. in prep.). The horizontal dashed lines denote the V-CDT standard isotopic ratios for \ce{^{32}S}/\ce{^{34}S}, \ce{^{32}S}/\ce{^{33}S}, \ce{^{34}S}/\ce{^{33}S} and \ce{^{16}O}/\ce{^{18}O} \citep{2001GeCoA..65.2433D, 2001RCMS...15..501W}.}
    \label{fig:fig4}
\end{figure*}

The isotope ratios of \ce{^{32}S}/\ce{^{34}S},  \ce{^{32}S}/\ce{^{33}S}, \ce{^{34}S}/\ce{^{33}S} and \ce{^{16}O}/\ce{^{18}O} measured from SO and \ce{SO_2} as a function of the assumed excitation temperature are shown in Figure~\ref{fig:fig4}. 

\end{document}